\documentclass[%
 reprint,
superscriptaddress,
 amsmath,amssymb,
 aps,
 floatfix,
 prd,
 nofootinbib,
]{revtex4-2}

\usepackage{graphicx}
\usepackage{dcolumn}
\usepackage{bm}
\usepackage[colorlinks=true, allcolors=blue, breaklinks=true, hyperindex=false]{hyperref}
\usepackage{xspace}
\usepackage{xcolor}

\newcommand{\Hunit}{\ensuremath{\text{km\,s}^{-1}\,\text{Mpc}^{-1}}}

\newcommand{\Hubble}{\ensuremath{H_{0}}\xspace}

\newcommand{\gwcosmo}{\texttt{gwcosmo}\xspace}

\newcommand{\mltp}{\textsc{MultiPeak}\xspace}
\newcommand{\fullpop}{\textsc{FullPop-4.0}\xspace}
\newcommand{\mdist}{Madau-Dickinson\xspace}
\newcommand{\gladeplus}{\textsc{GLADE+}\xspace}
\newcommand{\gwtc}{GWTC-4.0\xspace}
\newcommand{\pytorch}{\texttt{Pytorch}\xspace}
\newcommand{\oafour}{O4a\xspace}
\newcommand{\obfour}{O4b\xspace}
\newcommand{\ocfour}{O4c\xspace}
\newcommand{\ofive}{O5\xspace}
\newcommand{\params}{\ensuremath{\Lambda}\xspace}
\newcommand{\cosmoparams}{\ensuremath{\Lambda_{\text{cosmo}}}}
\newcommand{\popparams}{\ensuremath{\Lambda_{\text{pop}}}}
\newcommand{\ndet}{\ensuremath{N_{\text{det}}}\xspace}

\begin{document}

\preprint{APS/123-QED}

\title{Scalable Dark Siren Cosmology with \texttt{gwcosmo}: GPU Acceleration, Validation and Systematics}

\author{Alexander Papadopoulos}
 \affiliation{Institute for Gravitational Research, University of Glasgow, Glasgow, G12 8QQ, United Kingdom}

\author{Christian E. A. Chapman-Bird}
 \affiliation{Institute for Gravitational Wave Astronomy \& School of Physics and Astronomy, University of Birmingham, Edgbaston, Birmingham B15 2TT, United Kingdom}
 \affiliation{Institute for Gravitational Research, University of Glasgow, Glasgow, G12 8QQ, United Kingdom}

\author{Rachel Gray}
\author{Christopher Messenger}
 \affiliation{Institute for Gravitational Research, University of Glasgow, Glasgow, G12 8QQ, United Kingdom}

\author{Tom Bertheas}
 \affiliation{Laboratoire des 2 Infinis - Toulouse (L2IT-IN2P3), Université de Toulouse, CNRS, F-31062 Toulouse Cedex 9, France}
 \affiliation{Laboratoire de Physique de l’Ecole Normale Supérieure (LPENS), ENS, Université PSL, CNRS, 75005 Paris, France}
\date{\today}

\begin{abstract}
As the number of confident gravitational-wave detections grows, population-level hierarchical analyses face increasing computational costs.
Dark-siren cosmological inference integrates over the localisation volume of each gravitational-wave source. To remain feasible without discarding information from the quieter but more numerous sources in the catalogue, significant efficiency improvements are vital for analysis pipelines.
In this work, we present an upgraded version of the cosmological inference pipeline \gwcosmo, which leverages vectorisation on graphics processing units to process the entire gravitational-wave catalogue in parallel with each iteration.
This new implementation achieves a speed-up of $\sim 10^3$ over the previous version, facilitating analyses of O5-like numbers of GW events on wall-clock timescales of hours.
Our results demonstrate the scalability of the \gwcosmo pipeline, specifically its ability to handle the increasing computational load of expanding event catalogues, positioning it as a vital tool for future advances in dark-siren cosmology.

\end{abstract}

\maketitle

\section{Introduction}
Inference of cosmological parameters from compact binary coalescences (CBCs) was first described by Schutz in 1986~\cite{schutz1986}, introducing the usage of gravitational wave (GW) signals as ``standard sirens". Unlike electromagnetic distance ladders, standard sirens offer an independent probe of cosmology, making it possible to constrain parameters such as the Hubble constant, $H_0$, directly from GW signals. Cosmological inference with GW signals requires information on the redshift of the source (z), as well as the luminosity distance ($d_L$)~\cite{schutz1986,holz2005,moresco2022}. GW observations directly measure the redshifted chirp mass $\mathcal{M}_z = \mathcal{M} (1+z)$, where $\mathcal{M}$ is the source-frame chirp mass, and the luminosity distance of the source, which depends on redshift through the cosmological model
\begin{equation}\label{eq:h(z)}
d_L(z) = (1+z)\, \int_0^z \frac{cdz'}{H(z')}.
\end{equation}
Here $H(z)=H_0\sqrt{\Omega_m(1+z)^3+\Omega_\Lambda}$, with $\Omega_{m,\Lambda}$ being the matter and dark energy density parameters assuming a $\Lambda$CDM cosmological model. Because GW signals encode information on the redshifted rather than source-frame masses, the source redshift cannot be determined directly from GW data alone\footnote{In the case of binary neutron star mergers, it is possible to break this degeneracy using the tidal information of the signal~\cite{messenger2012}.}, and additional information is required to break this degeneracy
\par
In this framework, GW signals as standard sirens can be split into a number of classes. In the presence of an electromagnetic (EM) counterpart, the redshift can be determined by the localisation of the counterpart inside a host galaxy - earning the name \textit{bright sirens}
~\cite{holz2005,nissanke2010,GW170817}. When no unique counterpart information is available, it is possible to break the degeneracy by either using information solely from the mass distribution of GW sources, or combining this with galaxy surveys to statistically determine the redshift of the sources - \textit{spectral}~\cite{taylor2012,farr2019,ezquiaga2021,ezquiaga2022,pierra2026} and \textit{dark sirens} respectively~\cite{schutz1986,chernoff1993,delpozzo2012,soaressantos2019,Palmese_2020,finke2021,mastrogiovanni2021,gwcosmo2022,icarogw2023,borghi2024}.
\par
Since the first detection in 2015, more than 200 GW events have been detected by the LIGO-Virgo-KAGRA Collaboration (LVK)~\cite{ligo,virgo,kagra} up to the end of the first part of the fourth observing run (\oafour), collected in the fourth Gravitational-Wave Transient Catalog (\gwtc)~\cite{gwtc1,gwtc2,gwtc3,gwtc4results}. It is anticipated that this number will continue to rise in the near future, with both \obfour and \ocfour data releases still to come, as well as future observing runs. This rapidly growing number of detections will be crucial for tightening cosmological constraints with GWs, since the precision of these measurements increases with the number of observed sources.
Dark and spectral sirens will play an important role in constraining cosmological parameters, as to date there has been only one instance of a confidently associated bright siren, GW170817~\cite{GW170817}.
\par
The cosmological inference pipeline \gwcosmo~\cite{gwcosmo2020,gwcosmo2022,gwcosmo2023} implements the \textit{dark siren} method via a hierarchical Bayesian framework to jointly infer cosmological and population properties of compact binary coalescences, additionally incorporating information from galaxy surveys. Other pipelines such as \texttt{icarogw}~\cite{mastrogiovanni2023,icarogw} and \texttt{CHIMERA}~\cite{borghi2024,tagliazucchi2025} employ similar hierarchical methods to perform cosmological inference with GWs. Previous versions of \gwcosmo have been used by the LVK collaboration to analyse
GW events and produce cosmological constraints, most recently in~\cite{gwtc4cosmo}. 
During cosmological inference, this version of \gwcosmo operated on the GW catalogue data in a highly serialised fashion, processing the pixelated sky-localisation area of each GW event one at a time, working pixel-by-pixel to construct the hierarchical likelihood. This means that the wall-time for analysis scales approximately as $\mathcal{O}(N_{\mathrm{events}}\times N_\mathrm{pix}\times N_\mathrm{samples})$, denoting the number of events, pixels and samples in each pixel respectively. Indeed, in~\cite{gwtc4cosmo}, analyses using 141 GW events took $\mathcal{O}$(weeks) to reach convergence even when parallelised over $32$ 
central processing units (CPUs). This presents a real challenge to the scalability of cosmological analyses with a growing catalog of GW events, which threatens to become intractable in the current framework.
\par
In this work, we present an updated version of \gwcosmo which is capable of scalable cosmological inference using a graphics processing unit (GPU), as well as retaining the original ability to parallelise over multiple CPUs. This update allows us to perform fully vectorised evaluations of \gwcosmo's hierarchical likelihood, significantly decreasing the wall-time taken to produce constraints on cosmological and GW population parameters. Even when using $\mathcal{O}(10^3)$ GW events, comparable to the numbers expected during the \ofive observing run, the updated code is capable of outperforming its predecessor by a factor of 1000. 
This enables a wide range of new studies in GW cosmology in a fraction of the time, while producing results fully consistent with previous versions of the code and supporting future analyses with substantially expanded GW event catalogs through additional performance optimizations.
\par
This paper is organised as follows. In \S~\ref{sec:method} we describe the methods used which allow the likelihood to be evaluated on a GPU, as well as new features added to the pipeline. In \S~\ref{sec:results} we demonstrate the consistency of the code with the previous version, and the ability to recover parameters from a simulated population of GW events. Additionally in this section we investigate the effects of a number of implementation choices on the final cosmological constraints. We also include a number of appendices describing the GW data used in our analyses (Appendix \ref{app:data}), validating our criterion for posterior agreement (Appendix \ref{app:kl}), comparing code performance on different devices and energy consumption (Appendix \ref{app:devices}).

\section{Methodology}\label{sec:method}
\subsection{Hierarchical Likelihood}
For a population of GW sources described by hyperparameters \params={\cosmoparams,\popparams}, where \cosmoparams (\popparams) describe the cosmological (population) model, the posterior on the hyperparameters given the GW data $\{x_{\text{GW}}\}$ from the \ndet detected events, conditioned on the detection criterion $\{D_{\text{GW}}\}$ is given by by~\cite{mandel2019,vitale2021}:
\begin{equation}\label{eq:hier}
    \begin{aligned}
    p(\params|\{x_{\text{GW}}\},\{D_{\text{GW}}\},I) &\propto p(\params|I) p(\ndet|\params,I) \\
    & \times \prod_{i}^{\ndet} p(x_{\text{GW}i}|D_{\text{GW}i}, \params,I),
    \end{aligned}
\end{equation}
where the term $ p(x_{\text{GW}i}|D_{\text{GW}i}, \params,I)$ is the likelihood from GW event $i$ given the binary detection parameter $D_{\text{GW}i} = 1$. The inclusion of $I$ is a catch-all for the remaining information, cosmological or otherwise, not expressed explicitly. Using Bayes' theorem and by explicit marginalization over the source frame parameters of individual GW events $\theta$ (masses, spins, redshift, inclination etc.), this can be expressed as:
\begin{equation}\label{eq:hier_factorised}
    \begin{aligned}
    p(\params|&\{x_{\text{GW}}\},\{D_{\text{GW}}\},I) 
    \propto p(\params|I)p(\ndet|\params,I) \\ & \times \alpha(\params)^{-\ndet}
   \prod_{i}^{\ndet}\int p(x_{\text{GW}i}|\theta,\params,I)p(\theta|\params,I)d\theta.
    \end{aligned}
\end{equation}
\vspace{5pt}
The term $\alpha(\params)$ is known as the \textit{selection function}, and is defined as
\begin{equation}\label{eq:selfunc}
\alpha(\params)=\int p(D_{\text{GW}}|\theta,\params,I)p(\theta|\params,I)d\theta,
\end{equation}
and is used to account for the finite detection horizon for GW events in the observed population, meaning that they represent an incomplete set of the true astrophysical population. As this term is identical for all events in the population, it can be removed from the product. While seemingly dependent on \params, by using a scale-free Jeffrey's prior on the expected rate of detected GW events, the cosmological dependence in $p(\ndet|\params)$ drops out~\cite{fishbach2018}. 
\par
In order to explicitly incorporate galaxy catalog information, Eq.~\ref{eq:hier_factorised} can be rewritten to marginalise over GW event sky location ($\Omega$) by dividing each event into $N_\mathrm{pix}$ equally-sized pixels, and then marginalising over $z$, as well as explicitly including the selection function in the denominator,
\begin{widetext}
\begin{equation}\label{eq:hier_pixels}
    \begin{aligned}
        p(\params|\{x_{\text{GW}}\},\{D_{\text{GW}}\},I) &\propto{} p(\params|I)p(\ndet|\params,I) \left[ \iint p(D_{\text{GW}}|z, \theta', \params,I) p(\theta'|\params,I) \sum_{j}^{N_{\text{pix}}} p(z|\Omega_j, \params,I) d\theta' dz \right]^{-\ndet} \\
        & \times \prod_{i}^{\ndet} \left[ \iint \sum_{j}^{N_{\text{pix}}} p(x_{\text{GW}i}|\Omega_j, z, \theta', \params,I) p(\theta'|\params,I) p(z|\Omega_j, \params,I) d\theta' dz \right],
    \end{aligned}
\end{equation}
\end{widetext}
where $\theta = \{z,\theta'\}$. We approximate in Eq.~\ref{eq:hier_pixels} that the probability of detection does not vary over the sky, and so the $p(D_{\text{GW}}|z, \theta', \params,I)$ term is removed from the sum over pixels\footnote{This is because the probability of detection is averaged over an observing run, and the rotation of the earth blurs much of the dependence on sky position. There remains a mild dependence on declination which is neglected in this framework, consistently with the previous version of \gwcosmo~\cite{gwcosmo2023}.}. Of particular interest is the likelihood term $p(x_{\text{GW}i}|\Omega_j, z, \theta', \params,I)$. This can be written in terms of samples to explicitly show the usage of a kernel density estimate (KDE) to construct a distribution of GW probability in each pixel,
\begin{equation}\label{eq:hier_pixels_kde}
     p(x_{\text{GW}i}|\Omega_j, z, \theta', \params,I)\propto 
 \sum_{k=1}^{N_{\mathrm{samps},j}} w_{ijk}
    K_{h_{ij}}(z - z_{ijk})
\end{equation}
Where $K_{h_{ij}}(\cdot)$ is a Gaussian kernel and bandwidth $h_{ij}$, and the weights of sample $k$ in pixel $j$ are defined as
\begin{equation}\label{eq:weights}
    w_{ijk} = \frac{p( z_{ijk},\theta'_{ijk}|\params,I)}{\pi_{\mathrm{PE}}(z_{ijk},\theta'_{ijk})},
\end{equation}
which removes assumed parameter estimation priors, and re-weighting by the cosmological and population models being analysed. Lastly, we also include a numerical stability criterion on a pixel-by-pixel basis called the \textit{effective number of samples}, $N_\mathrm{eff,PE}$, which for pixel $j$ in event $i$ using samples $k$ is defined as~\cite{farr2019accuracy,talbot2023}
\begin{equation}\label{eq:neff}
    N_{\mathrm{eff,PE},j} = \frac{(\sum_kw_{ijk})^2}{\sum_kw_{ijk}^2},
\end{equation}
using the weights calculated in Eq.~\ref{eq:weights}. If a pixel does not pass this threshold for a given set of \params, it does not contribute to to the total likelihood. For further discussion of the statistical framework underpinning \texttt{gwcosmo}, see~\cite{gwcosmo2023}.
\subsection{GPU Acceleration}
\begin{figure}[h]
    \centering
    \includegraphics[width=\linewidth]{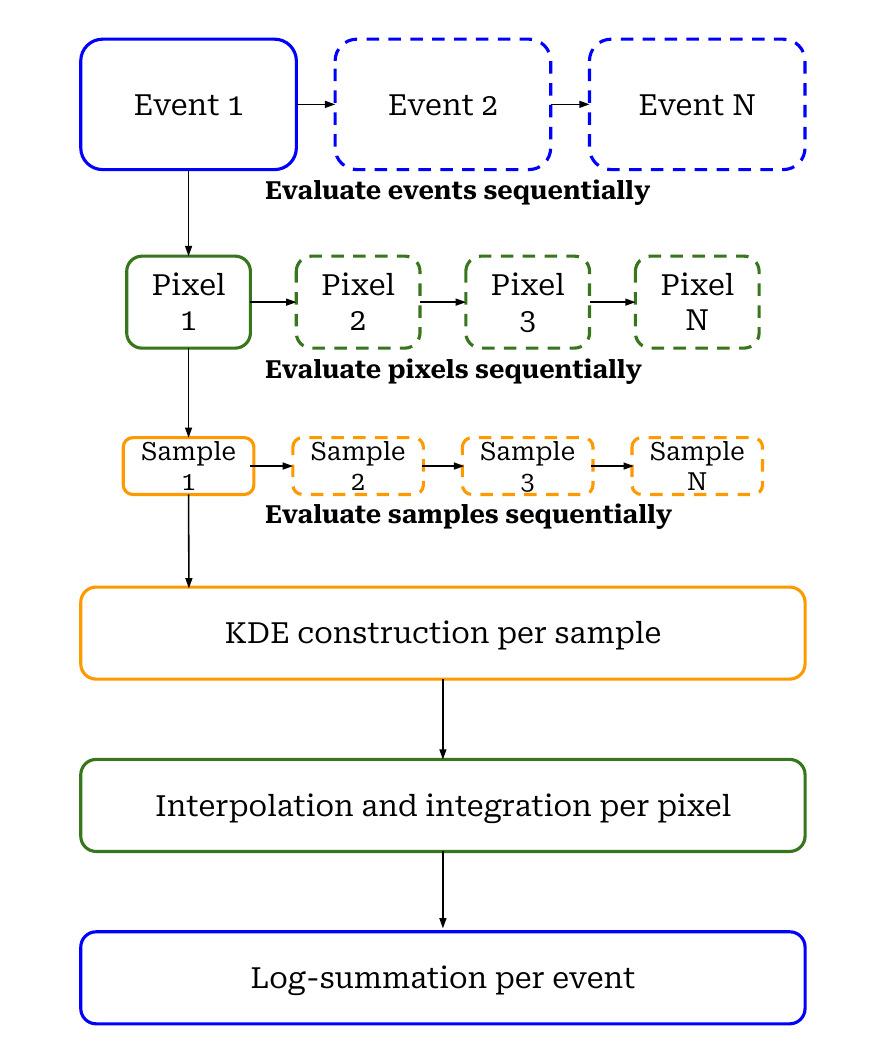}
    \caption{Sequential CPU based likelihood evaluation scheme. The process uses serial evaluation over events, pixels and samples, whereby the KDE construction iterates over each sample, then pixel-level distributions are interpolated and integrated, and the single-event likelihoods are log-summed. Dashed lines indicate elements which must wait to be evaluated in the sequential scheme.}
    \label{fig:cpuchart}
\end{figure}
\begin{figure}[h]
    \centering
    \includegraphics[width=\linewidth]{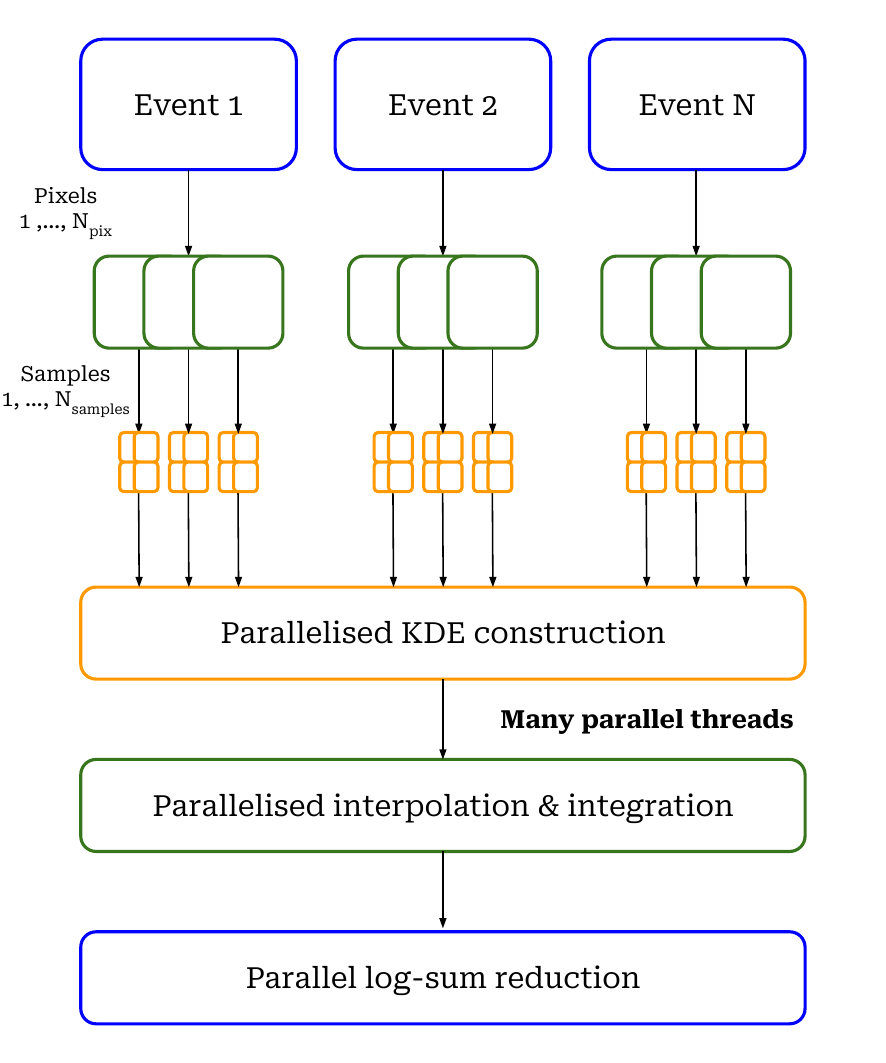}
    \caption{Parallel GPU based likelihood evaluation scheme. The strength of this implementation is to leverage many parallel threads in order to simultaneous process events, pixels and samples. The parallelised KDE construction, interpolation and integration, and log-sum reduction work in tandem to produce a significant performance upgrade relative to the sequential implementation.}
    \label{fig:gpuchart}
\end{figure}
The primary computational cost when evaluating the \gwcosmo's likelihood is constructing the KDEs in Eq.~\ref{eq:hier_pixels_kde}. These are evaluated over the redshift samples - which are dependent on \params via \cosmoparams - for all sky pixels in each GW event. For each likelihood evaluation \params are updated, and so the KDEs must be re-computed on the fly for each event. Given the lack of an alternative method to compute these in parallel, this operation is sequential, meaning the cost increases linearly with the number of events in the analysis. This process, which was implemented in previous iterations of \gwcosmo , is visualised in Fig.~\ref{fig:cpuchart}. In each likelihood call, the code must move through a set of nested loops, with the top level being a loop over $N_\mathrm{events}$, then through each of the $N_\mathrm{pixels}$ making up the GW event sky area, and finally through the $N_\mathrm{samples}$ in each pixel when constructing the KDE. These pixels are then integrated over redshift, before a summation on the event level, to produce a single hierarchical likelihood value. The cost of this process grows linearly as the number of GW events being analysed increases, to the point that analyses become intractable.
\par 
We present a new approach to evaluate the \gwcosmo likelihood, which leverages the massively parallel processing capabilities of GPUs to significantly improve the performance of cosmological analyses. The major structural change required to implement this is the construction of a three-dimensional tensor that contains event posterior samples, of shape 
($N_{\text{events}},N_{\text{pix}},N_{\text{samples}}$). Each tensor element corresponds to an unweighted posterior sample associated with a given sky pixel for a given event. To avoid the complexity of ragged arrays required by events with different numbers of pixels and samples, we pad our tensors based on the largest member's dimensions (max sky pixels and max samples). This ensures a dense, rectangular structure where padded indices do not contribute anything to the total likelihood. As visualised in Fig.~\ref{fig:gpuchart}, this allows the likelihood evaluation to be fully vectorized, with every operation contained in both numerator and denominator of Eq.~\ref{eq:hier_pixels} being computed in parallel over events, pixels and samples. This means that all explicit loops over events, pixels and samples are removed from the code, delivering a significant increase in computational performance. The cost of this process is in theory $\mathcal{O}(1)$, i.e constant with respect to the size of the inputs, but in practice this will be limited by both the part of the likelihood which is not computed in parallel, and by the total GPU memory available. 
\par 
To allow for this, we implement a \pytorch~\cite{paszke2019} framework for \gwcosmo, making population models, cosmological calculations and selection effects computation \pytorch-compatible. Additionally, a number of custom \texttt{CUDA}~\cite{cuda} kernels are implemented using the \texttt{Numba}~\cite{numba} library. In particular, the vectorized computation of KDEs eliminate the computational bottleneck caused by sequential KDEs. These kernels allow for many operations to be carried out at low-level using shared memory, removing the need to instantiate large intermediate arrays during the likelihood evaluation. Despite these substantial changes to allow the code to run in a massively parallel framework, it retains the CPU-only capability of previous versions - ensuring backwards compatibility.

\subsection{Memory Reduction Strategies}\label{subsec:memory}
The structure of a three-dimensional rectangular array containing the GW event posterior samples is very effective in improving the performance of the pipeline, however this is a trade off in terms of the memory required. Given that GW posterior samples can number in the thousands for a given event, and the size of one array dimension is set by the maximum amount in \textit{any} pixel in \textit{any} event, the memory requirement of analysis can balloon with only a small number of events. In order to combat this, we implement the option to randomly downsample the GW posterior samples in a given pixel. In the limit of many thousands of samples, this still allows the KDE to accurately represent the distribution in each pixel. 
\par
Additionally, for a catalog of many thousands of events, we can directly decrease the size of the array being evaluated by batching the likelihood. This is particularly useful when using a device with less VRAM, but also sees a decrease in computational performance proportional to the fraction of events evaluated in each batch.

\section{Results}\label{sec:results}
\subsection{Acceleration}
To demonstrate the improved computational performance of \gwcosmo, we benchmark the code using 2000 simulated GW events, more than 10 times the maximum number used in an analysis to date. The generation of this data is described in Appendix~\ref{app:data}. We compare the GPU implementation to the CPU code used in~\cite{gwtc4cosmo}, and all values are produced with an Nvidia H100 GPU using Hopper architecture unless explicitly stated otherwise. The CPU values presented here were produced using 2.4GHz Xeon E5-2630v3 CPUs, with the results in Fig.~\ref{fig:gwtc4_corner} parallelised over 32 processes. When stating times we take the mean evaluation time over 100 evaluations of the likelihood for hyperparameters randomly drawn from the prior, after 10 warm-up evaluations which are discarded. 
\par
\begin{figure}
    \centering
    \includegraphics[width=\linewidth]{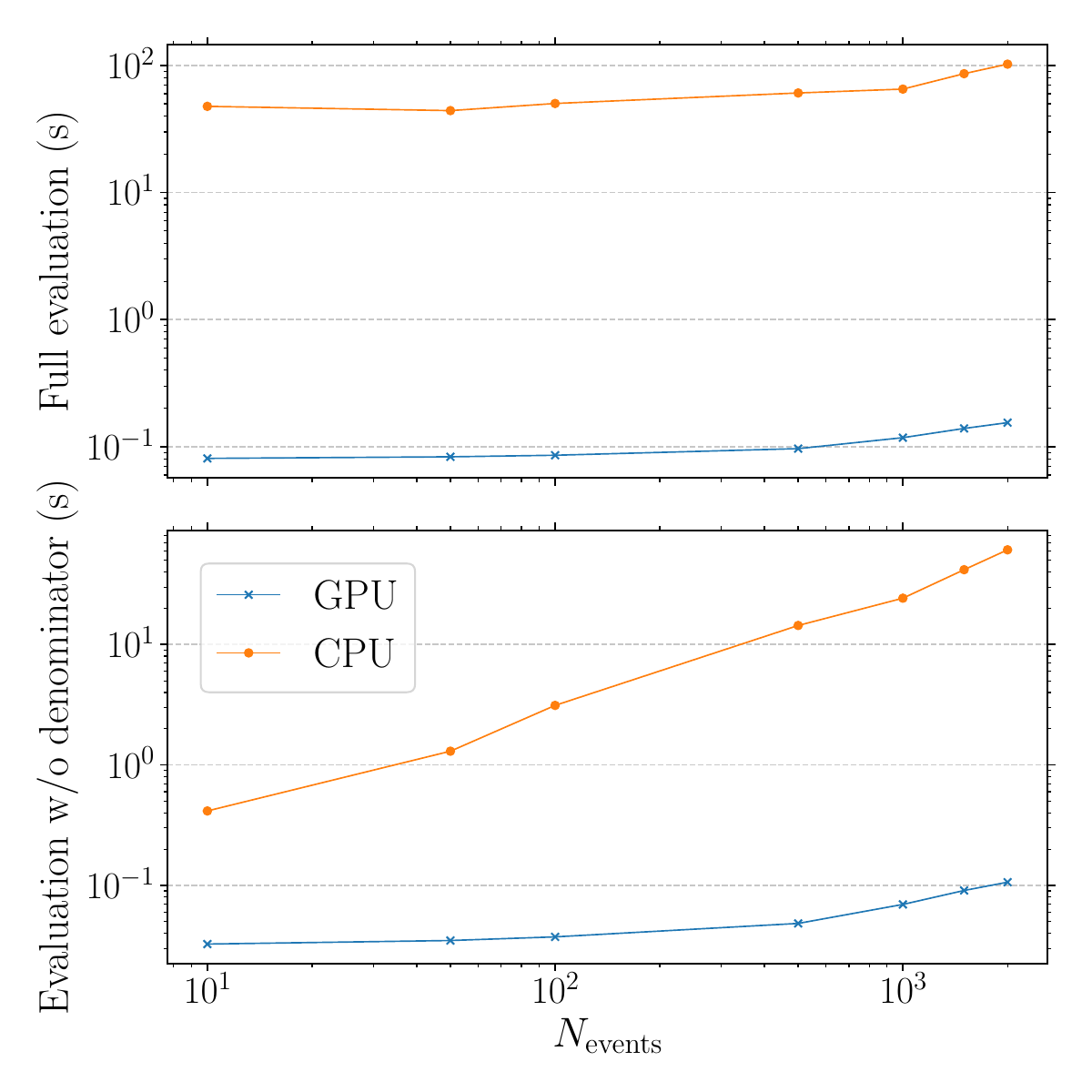}
    \caption{Mean single likelihood evaluations times for the CPU and GPU implementations as a function of events, averaged over 100 evaluations. Both curves use the full set of GW event samples for a direct comparison. GPU values are calculated using an Nvidia H100 GPU. \textit{Top panel}: Evaluations of both the numerator and denominator of the likelihood. For the legacy CPU implementation the evaluation of selection effects becomes a dominant cost. \textit{Bottom panel}: Evaluations of the numerator only, where the linear scaling of the legacy CPU implementation is clear.}
    \label{fig:cpuvsgpu}
\end{figure}

In Fig.~\ref{fig:cpuvsgpu} we compare the two versions of \gwcosmo as a function of the number of events in the catalog, ranging from $10$ to $2000$. The top panel includes the total evaluation time (numerator + denominator + parameter updates, from Eq.~\ref{eq:hier_pixels}), while the bottom panel is the full likelihood without the denominator included. This was motivated by the fact that for O5-like sensitivity~\cite{o5like}, a very large number of injections are required to calculate selection effects - in the previous version of \gwcosmo this becomes a dominant cost. This is due to the larger parameter volume which becomes accessible to the detector - in order to accurately calculate Eq.~\ref{eq:selfunc} we require many more injections. When excluding this, the CPU evaluation curve returns to an expected linear scaling with $N_\mathrm{events}$. The key takeaway is the improvement of a factor of almost 1000 when $N_\mathrm{events}=2000$, ensuring the scalability of \gwcosmo for \ofive-like analyses. Although in theory a perfectly vectorised implementation should not scale at all with $N_\mathrm{events}$, the small increase in computational time taken on the GPU can be ascribed to a small number of operations which still require iterating over events, pixels or samples - as opposed to GPU saturation.
\par
\begin{figure}
    \centering
    \includegraphics[width=\linewidth]{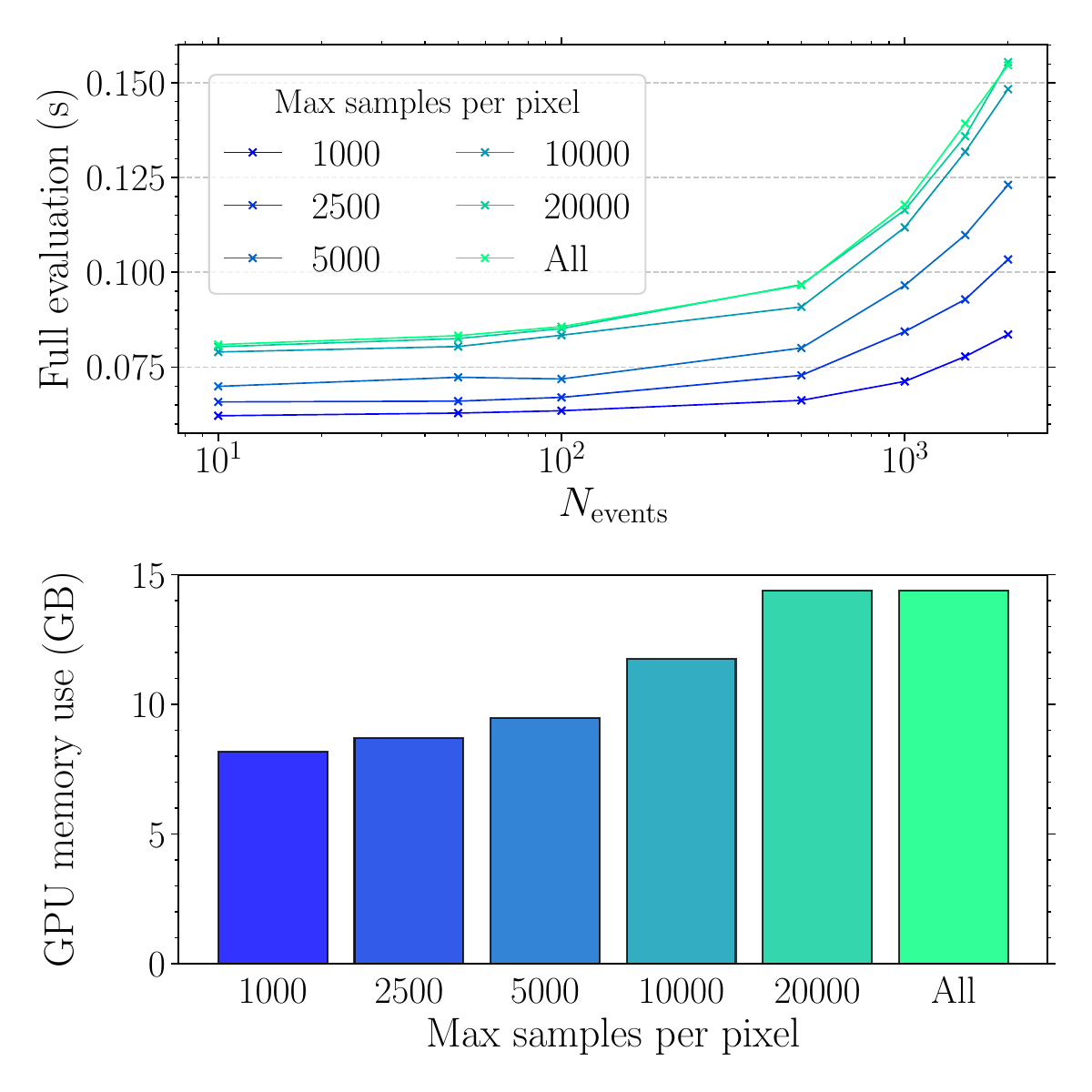}
    \caption{\textit{Top panel}: Likelihood evaluation times for the GPU code as a function of GW event catalog size, with different pixel downsampling limits, averaged over 100 evaluations. By decreasing the maximum number of samples used the mean evaluation time can be decrease by a further factor of $2$ for the largest catalog size. \textit{Bottom panel}: Maximum memory usage as a function of downsampling limits when using a full event catalog of 2000 GW sources. Event a conservative downsampling can significantly reduce the total memory required by the code. These values were produced enforcing a minimum number of pixels per event of $5$.}
    \label{fig:perform}
\end{figure}

It is possible to further decrease the computational cost of evaluating the likelihood by using a subset of the total posterior samples available per pixel, as described in \S~\ref{subsec:memory}. In Fig.~\ref{fig:perform} we see that by decreasing this cap (effectively also a cap on the total number of samples), an improvement of a factor of $2$ in performance can be achieved. In the lower panel of the same figure, the potential memory-saving effects of this are also clear. In \S~\ref{subsec:systematics} we investigate the effects of this on the final posterior. 
\par
In Table~\ref{tab:profiling} we break down the cost of a single likelihood evaluation by component and compare the improved performance relative to the legacy CPU implementation. We break evaluation into the numerator, denominator and other data processing operations, and compared these for the CPU, and two GPU settings - fiducial (with all samples) and downsampled (using a cap of $2500$ samples per pixel). The main operations in the numerator are the construction of KDEs and interpolating and summing over pixels, before integrating over the expression. These operations are extremely costly in the CPU implementation, making up over half the total evaluation time. Even for the fiducial GPU case, the KDE is a factor of almost $400\times$ faster, while the interpolation and summation are sped up by a factor of greater than $3\times10^4$. In the CPU implementation, the denominator becomes a more significant computational bottleneck (independent of the GW catalog size). When considering detectors with higher sensitivities, as a greater number of injections are required to evaluate the selection effects. The effect of this is clear in the CPU column of Table~\ref{tab:profiling}. By instead handling this with a \pytorch framework the cost is reduced by a factor of almost $900$. For more straightforward operations on flattened arrays such as reweighting of GW event samples or filling arrays, the GPU implementation actually increases the cost due to additional overhead, however the effect of this on the total evaluation time is negligible due to enormous savings in other operations. Overall, the improvement in performance is stark - with even the most conservative settings provided inference accelerated by a factor of greater than $600$. In addition to this, by introducing mild downsampling to the posterior samples in each pixel can result in a further acceleration, meaning a single likelihood evaluation is complete more than one thousand times quicker than with the CPU implementation.

\begin{table*}
\centering
\begin{tabular}{lr@{\quad}rr@{\quad}rr}
\hline
\textbf{Component} & \textbf{CPU} & \multicolumn{2}{c}{\textbf{GPU Fiducial}} & \multicolumn{2}{c}{\textbf{GPU Downsampled}} \\
 & \textbf{Time (ms)} & \textbf{Time (ms)} & \textbf{Speedup} & \textbf{Time (ms)} & \textbf{Speedup} \\
\hline
\multicolumn{6}{l}{\textit{Numerator}} \\
\quad KDE & 29780.13 & 75.59 & 394.0$\times$ & 28.22 & 1055.3$\times$ \\
\quad Interpolation \& Rate & 32156.28 & 1.04 & 30919.5$\times$ & 1.02 & 31525.8$\times$ \\
\quad Integration & 200.45 & 0.19 & 1055.0$\times$ & 0.19 & 1055.0$\times$ \\
\hline
\multicolumn{6}{l}{\textit{Denominator}} \\
\quad Selection Effect & 41772.22 & 48.70 & 857.7$\times$ & 48.76 & 856.7$\times$ \\
\quad $z$ Prior & $<0.01$ & 0.53 & 0.33$\times$ & 0.52 & 0.33$\times$ \\
\hline
\multicolumn{6}{l}{\textit{Other Operations}} \\
\quad Sample Reweighting & $<0.01$ & 22.55 & 0.00005$\times$ & 14.69 & 0.00008$\times$ \\
\quad Parameter Update & 0.81 & 6.61 & 0.12$\times$ & 6.43 & 0.13$\times$ \\
\quad Fill Arrays & $<0.01$ & 5.38 & 0.00004$\times$ & 1.88 & 0.0001$\times$ \\
\quad Redshift Rate & 0.12 & 0.11 & 1.10$\times$ & 0.11 & 1.09$\times$ \\
\hline
\textbf{Total} & \textbf{103910.31} & \textbf{160.70} & \textbf{646.6$\times$} & \textbf{101.82} & \textbf{1020.5$\times$} \\
\hline
\end{tabular}
\caption{Comparison of the evaluation cost for the legacy CPU likelihood and the GPU implementation, broken down by component when using a catalog of 2000 GW events. The `CPU' column profiles the legacy implementation, while `GPU Fiducial' reports the GPU likelihood using all PE samples, and `GPU Downsampled' reports the GPU likelihood with a maximum of $2500$ samples per pixel. The KDE evaluation, interpolation and selection effect calculation dominate the total time in the CPU implementation, and are significantly accelerated by the GPU implementation.}
\label{tab:profiling}

\end{table*}

\subsection{Code validation}
In order to verify the consistency of the GPU implementation of \gwcosmo, we report two tests: firstly, recovering the injected hyperparameters of a simulated population of $2000$ GW events - representing an \ofive-like magnitude - in order to demonstrate the self-consistency of our method. The generation of this simulation population is described in Appendix~\ref{app:data}. Secondly, we analyse 141 real GW events from \gwtc and look for agreement with the same analysis produced by the previous CPU-only implementation of the framework. For all analyses presented in this section we employ the \texttt{bilby} parameter estimation library, with the nested sampler \texttt{nessai}~\cite{bilby_paper,nessai,nessai2023}.
\par
\begin{figure*}
    \centering
    \includegraphics[width=0.85\linewidth]{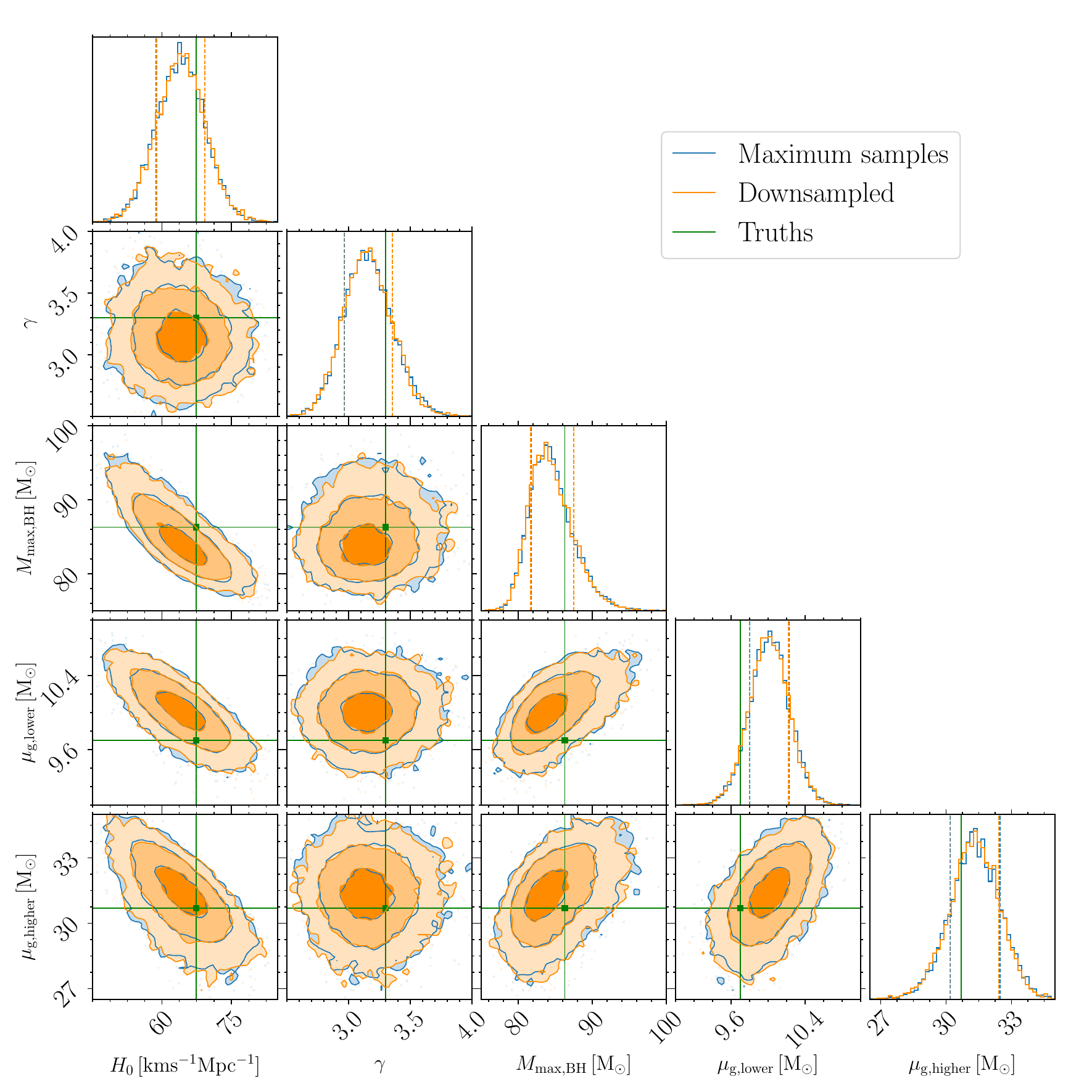}
    \caption{Corner plot of recovered hyperparameters using a simulated set of $2000$ GW events at \ofive sensitivity using the \mltp mass model, analysed with the GPU version. A subset of the total 15 hyperparameters are included here, of which the most cosmologically informative are chosen along with $H_0$. The blue (orange) posteriors pertain to analysis using all (2500) samples allowed in each pixel. Both sets of posteriors are in good agreement and recover the injected  values, shown by the green lines.}
    \label{fig:mock_data_corner}
\end{figure*}

 
Figure~\ref{fig:mock_data_corner} presents the results of our analysis with simulated data for two different settings, with all samples being used and when capping the maximum number allowed per sky pixel. Using 2000 mock GW events in a spectral siren analysis (without galaxy catalog information), we find that the new code version correctly recovers the injected hyperparameters for both cases, with the posteriors in good agreement. It also projects that such a spectral analysis with the BBH-only\mltp mass model produces a constraint of $\Hubble=64.01^{+5.20}_{-5.30}\, \Hunit$, or with an uncertainty of $16\%$. This analysis with an O5-like event count was able to reach convergence in a sampling time of 21 hours on a GPU when using the maximum available number of event samples per pixel, and when capping this at 2500 samples, of 17 hours. This cap on maximum samples per pixel is motivated in \S~\ref{subsec:systematics}. It is worth noting that this analysis would be infeasible wit the previous CPU version of \gwcosmo, due to to the total computational cost.
\par
\begin{figure*}
    \centering
    \includegraphics[width=0.85\linewidth]{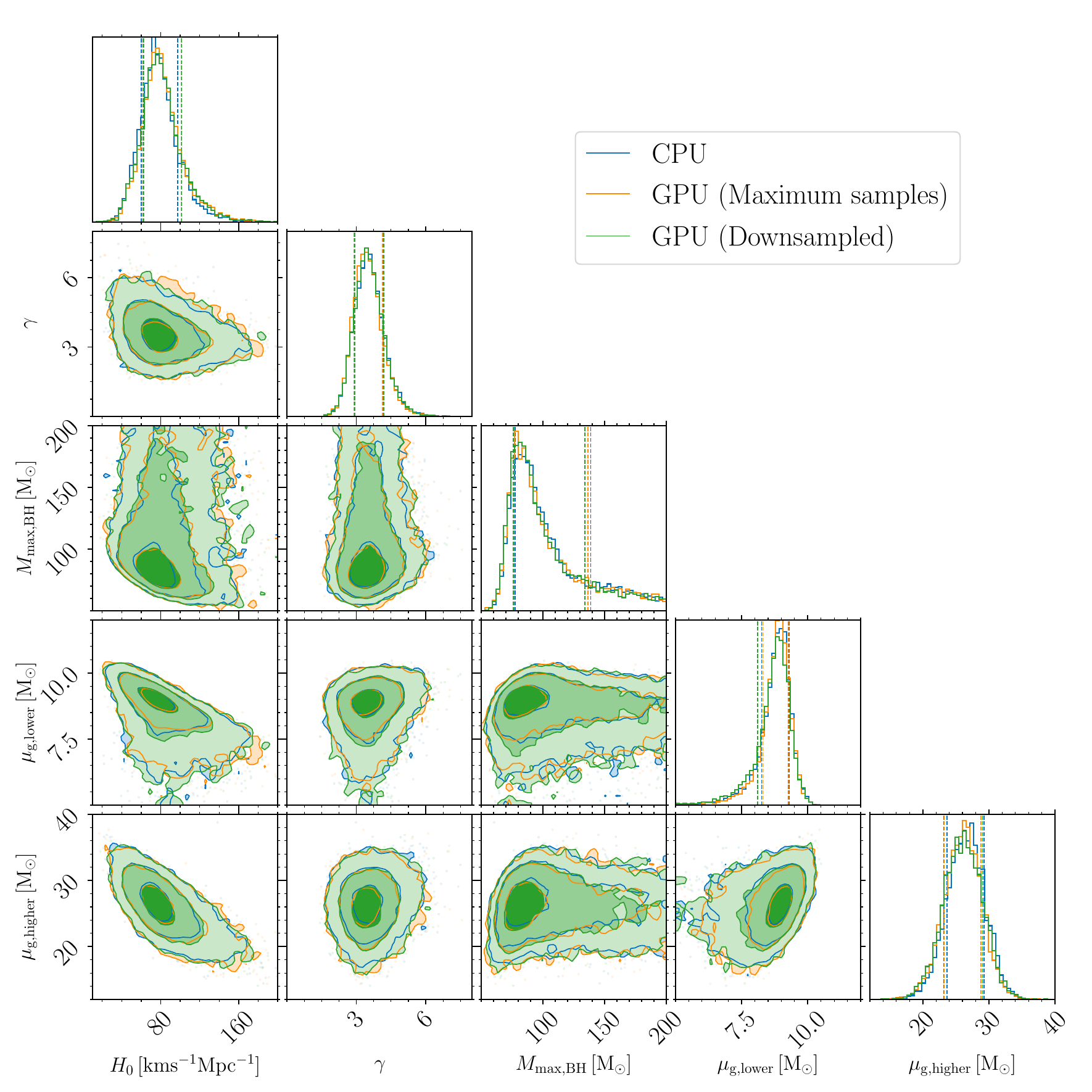}
    \caption{Comparison of posteriors produced using the legacy CPU implementation and the GPU implementation of \texttt{gwcosmo} for a dark siren analysis with \gwtc using the \fullpop mass model. The blue posteriors represent the CPU analysis, while the orange (green) posteriors show analysis using all (2500) allowed samples per pixel. The posteriors demonstrate excellent agreement between all three analyses, with the GPU analysis using all samples. We determine agreement between the posteriors by calculating a bootstrap noise threshold using the CPU posterior (described in Appendix~\ref{app:kl}), which we find to be $118.7\,\mathrm{mbits}$. We find that in comparison with this, maximum marginal KL values for the GPU ($96.0\,\mathrm{mbits}$) and GPU downsampled ($86.4\,\mathrm{mbits}$) both fall beneath this, indicating agreement between the posteriors. These three analyses converged in 768 hours, 96 hours and 15 hours for the CPU, GPU and downsampled GPU results respectively.}
    \label{fig:gwtc4_corner}
\end{figure*}

To demonstrate the ability of the updated code to produce posteriors in agreement with previous iterations, we compare identical dark siren analyses using \gwtc events using the \fullpop~\cite{gwtc4rp,gwtc4cosmo} model. This is presented in Figure.~\ref{fig:gwtc4_corner}, with the legacy CPU result in blue and two GPU results with different maximum sample caps in orange and green. These posteriors demonstrate excellent agreement, which is quantified using the Kullback-Leibler (KL) divergence between the marginal posteriors of each parameter~\cite{kldiv}. The GPU posteriors are considered in agreement with the CPU reference when the 90th percentile of marginal KL divergences falls below a noise tolerance bound established from the CPU distribution itself. This process is described in more detail in Appendix~\ref{app:kl}. For the full GPU analysis, the 90th percentile marginal KL is $94.0\,\mathrm{mbits}$, compared to a noise bound of $118.7\,\mathrm{mbits}$, confirming agreement. The downsampled GPU analysis achieves a 90th percentile KL of $86.4\,\mathrm{mbits}$, demonstrating that the approximation does not 
degrade the inference.
\par
Beyond this, the improvement in performance between the two implementations is significant - the GPU using all samples analysis reached convergence in 4 days (96 hours) of sampling, compared to 32 days (768 hours, or 24,576 hours core-hours distributed over 32 cores) for the legacy CPU analysis. When decreasing the maximum number of samples per pixel to 2500, this is further improved to just 15 hours of sampling time - highlighting the performance gain from this approximation, while also showing that it does not degrade the inference. It is also worth noting that this comparison does not take into account the CPU analysis being parallelised over 32 cores, despite this GPU analysis is significantly faster in both cases. In Appendix.~\ref{app:devices} we also note that the energy consumption of the two different implementations is very different, with the CPU-bound inference requiring an order of magnitude more energy to reach convergence than the GPU implementation using downsampling.
\par
These analyses cannot be directly compared to the time required in our mock data analyses however, due to the presence of an additional 8 hyperparameters in the \fullpop mass model relative to the \mltp, as well as the \fullpop model being more complex and taking longer to evaluate for a given set of \popparams. For more details on the construction of the two models, see~\cite{gwtc4cosmo}.
\subsection{Systematic tests}\label{subsec:systematics}
The improved performance of \gwcosmo enabled a systematic validation of several implementation choices. Five configuration settings were varied to assess their impact on the resulting $H_0$ posterior: the previously introduced cap on the maximum number of samples per sky pixel, the bandwidth selection method for the sky-pixel KDEs, the resolution of the KDE evaluation grid, the floating-point precision, and the minimum effective number of PE samples required per pixel. Each case was compared to the "fiducial" analysis settings, using all available GW event samples, single precision floats, the densest KDE grid, the default bandwidth estimator and $N_{\mathrm{eff,PE}}=2$ as defined in Eq.~\ref{eq:neff}.
\par
\begin{figure*}
    \centering
    \includegraphics[width=0.85\linewidth]{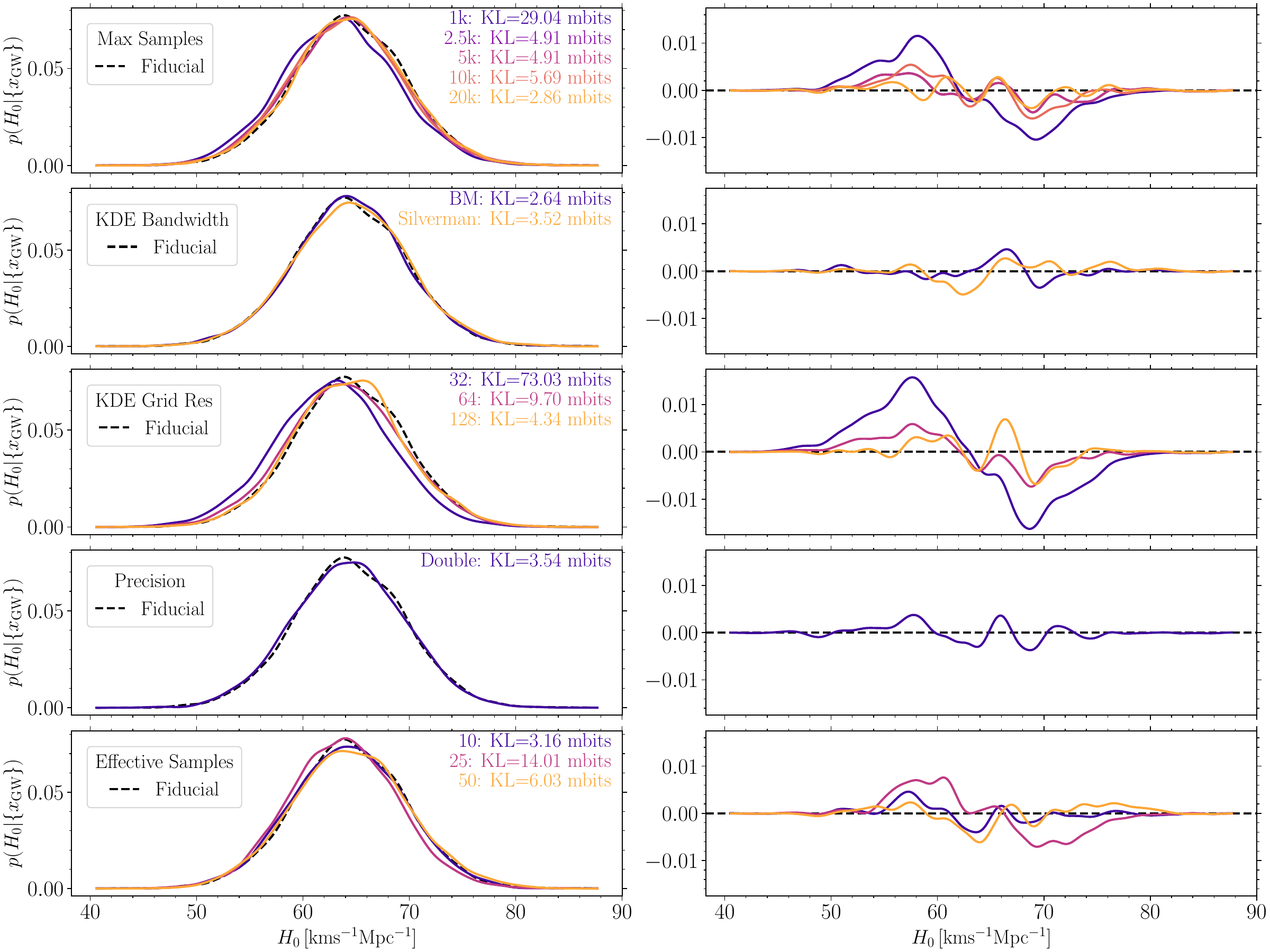}
    \caption{\textit{Left panel}: $H_0$ posteriors for systematic test cases, in descending order: maximum allowed samples per pixel, KDE bandwidth estimator, KDE grid resolution, floating point precision, minimum number of effective samples. The fiducial case is shown with the dashed black line, against which the KL divergence are calculated, whilst different tests are shown in different colours \textit{Right panel}: Residuals for $H_0$ posteriors for systematic test cases, calculated against the fiducial case represented by the black dashed line.}
    \label{fig:systematics_gallery}
\end{figure*}

The left panel of Fig.~\ref{fig:systematics_gallery} shows the resulting $H_0$ posteriors for each test, as well as the KL divergence between the fiducial and test cases. We determine that a KL value below or equal to $5.96 \,\mathrm{mbits}$ for the $H_0$ posterior indicates agreement, a choice which is motivated in Appendix~\ref{app:kl}. Based on this criterion, we find that it is possible to place a cap of $2500$ samples per sky pixel without seeing a significant shift in the final $H_0$ posterior or losing any constraining power relative to an analysis using all GW event samples. Below this value however we find a difference in the posterior no longer consistent with noise from a finite number of samples being use to calculate the KL divergence. The maximum sample cap does give an additional improvement in performance, with a run using a cap of $2500$ samples converging in $\sim 15$ hours, compared to $\sim 21$ hours when using the complete set of GW event samples. There is a slight systematic shift as the maximum number of samples decreases. This occurs because uniform downsampling inside a pixel alters the representation of the asymmetric sample distributions typical of GW distance posteriors. The resulting loss of resolution in the KDE tails ultimately affects the shape and position of the $H_0$ posterior. However, this effect causes only a small difference in the overall posterior and still falls below our criterion for agreement.
\par
We also find that calculating KDE bandwidths with more conservative estimators does not significantly affect the resulting posteriors on $H_0$. The default estimator uses Scott's rule of thumb~\cite{scott}, but we additionally consider Silverman's rule of thumb~\cite{silverman},and the biweight-midvariance (BM)~\cite{biweight}. These were chosen to be more robust to potential outliers in the sample set, and are outlined in more detail in Appendix~\ref{app:ests}. Despite this, neither has a significant effect on the posterior, nor do they affect the total convergence time for analysis. 
\par
A greater effect is found by decreasing the resolution of the KDE evaluation redshift grid from the fiducial $256$ points. This is the number of points over which the KDE is constructed, and by increasing or decreasing it the ability of the KDE to recreate structure is improved or worsened. As seen in the centre row of Fig.~\ref{fig:systematics_gallery}, when this is below $128$ points the posterior shifts relative to the fiducial case and no longer meets our KL divergence criterion. This can be explained by the fact that the grid is no longer dense enough to correctly capture the structure in the GW event samples. Indeed, this is one area in which the new version of \gwcosmo improves over it's predecessor, which only used $100$ points for the KDE evaluation grid due to linear computational cost. This obstacle is removed in the new, vectorised code and as such finer grids can be used.
\par
The results also demonstrate that using single precision floating point numbers shows no disadvantages relative to double precision. Indeed, the final posteriors meet our criterion for agreement, while the double precision analysis took over double the total analysis time compared to single precision, with $\sim46$ hours compared to $\sim$ 21 hours. This test allows us to state that our analyses are unchanged by the use of single single precision floats.
\par
Lastly, in the bottom left panel of Fig.~\ref{fig:systematics_gallery}, we test the effect of numerical stability criteria on our analysis. Each pixel in the likelihood must pass a threshold on $N_\mathrm{eff,PE}$, as defined in~\ref{eq:neff}. If a pixel does not pass the threshold on $N_\mathrm{eff,PE}$ then it does not contribute to the likelihood. This threshold is also dependent on the minimum number of pixels enforced per event, as having fewer pixels with more samples means each is more likely to pass the threshold. However we do not vary this in this work, so that the systematic tests presented here remain consistent. We find that imposing a harsher threshold on the number of effective GW event samples in the bottom row of the left panel of Fig.~\ref{fig:systematics_gallery} does show deviations in the posterior. Increasing this above $N_\mathrm{eff,PE}=10$ introduces more variance to the shape of the posterior and it no longer passes our agreement criterion.
\section{Conclusions}
We have developed a new implementation of \gwcosmo, which takes advantage of the massively-parallel computational ability of GPUs, by enabling near-fully vectorised evaluation of the likelihood over GW events, pixels and samples. By restructuring the hierarchical likelihood and using a dense tensor containing the GW data, as well as implementing \pytorch functionality and custom \texttt{CUDA} kernels for key operations, we have eliminated the dominant sequential bottlenecks which were present in previous CPU-only versions of \gwcosmo. Beyond this, it also retains its predecessor's CPU capability, to allow analysis in different computing environments. 
\par
This results in a potential performance increase of $\mathcal{O}(10^3)$, reducing likelihood evaluation times from minutes to milliseconds for larger gravitational wave event catalogs. Given this, we demonstrated the ability of \gwcosmo to analyse O5-like numbers of GW events ($\mathcal{O}(10^3)$) in significantly reduced time, making previous intractable analyses computationally feasible. The GPU implementation is also capable of analysing substantially larger catalogs, by using the batched likelihood as described in Sec.~\ref{subsec:memory}, at the cost of computational performance. In order to further optimise the memory efficiency of the code, and allow analyses of larger event catalogues, it will be necessary to eliminate padded arrays and replace them with flattened contiguous memory layouts and offset-based ragged tensor representations, reducing wasted memory and improving GPU utilisation
\par
With this new implementation, we demonstrated consistency of the pipeline through both the recovery of injected hyperparameters from a simulated population of GW events, as well as agreement with posteriors produced with real data using the legacy CPU implementation. As shown in Appendix~\ref{app:devices} this also provides much improved energy efficiency for cosmological inference, with a single analysis requiring an order of magnitude less energy when using a GPU. 
\par
Our testing also reveals that single-precision floating point arithmetic, or moderate downsampling of GW event samples do not introduce significant deviations in the final posteriors. Both of these are valuable tools with two purposes, allowing for mitigation of out-of-memory issues, as well as additional performance increases. However a decrease in the resolution of the KDE grid or an aggressive stability cut on sky pixels have a more significant effect on the final posterior. 
\par
The new vectorised implementation of \gwcosmo also makes it competitive in performance with other cosmological inference pipelines. For example, \texttt{icarogw}~\cite{icarogw} uses Monte Carlo integration rather than KDEs to evaluate GW posterior samples, yet achieves comparable runtimes. In the case of \texttt{CHIMERA}~\cite{borghi2024,tagliazucchi2025}, the ``many-1D" kernel which is most similar to that in \gwcosmo has a similar evaluation time for O5-like event numbers. An important caveat for these comparisons is the use of different GPU devices, and so they are not a direct comparison.
\par
Overall, this implementation will enable a new regime of dark siren cosmology using \gwcosmo, ensuring cosmological inference remains computationally tractable, even in an era of ever-expanding GW event catalogs. This capability opens the door to more sophisticated population modelling, the incorporation of new galaxy catalogs (see for example~\cite{McMahon:2026nhi}, which made use of the version of \gwcosmo presented in this paper), and extended studies of modified gravitational-wave propagation.

\begin{acknowledgments}
A.~Papadopoulos is supported by UKRI STFC studentship 323353-01. C. E. A.~Chapman-Bird. is supported by UKSA
grant UKRI971. R.~Gray and C.~Messenger are supported by the University of Glasgow and STFC grant ST/V005634/1.  T.~Bertheas acknowledges support from a CDSN PhD grant from ENS-PSL. This material is based upon work supported by NSF's LIGO Laboratory which is a major facility fully funded by the National Science Foundation. The authors are grateful for computational resources provided by the LIGO Laboratory and supported by National Science Foundation Grants PHY-0757058 and PHY-0823459, as well as by the IN2P3 computing centre (CC-IN2P3) in Lyon (Villeurbanne).
\end{acknowledgments}
\appendix
\section{Data}\label{app:data}
\textit{Simulated Data}:
To test and benchmark the performance of \gwcosmo, we simulate a catalog of mock gravitational wave events at \ofive sensitivity~\cite{o5like}. These were generated assuming a realistic 4-detector LVK network complete with 70\% uncorrelated duty cycles, and PE carried out using \texttt{bilby} with the \texttt{IMRPhenomXHM} waveform and gaussian noise. The population of GW events was drawn from the \mltp mass model, covering only the BBH range of the CBC mass spectrum, as first used in~\cite{gwtc2rp} and assuming a \mdist distribution in redshift~\cite{mddist}. We generate 2000 GW event posteriors with an SNR $>\,12$, which approximates the number expected for three years of observing at \ofive design sensitivity. This is not intended to be a projection of the potential constraints on $H_0$ in this regime, but to demonstrate that \gwcosmo can analyse such a magnitude of events. 
\begin{table}[h]
\centering
\begin{tabular}{lc@{\hspace{1.5em}}lc}
\hline
Parameter & Value &
Parameter & Value \\
\hline
$\alpha$                             & 2.9 &
$\mu_{g,\rm high}\,[M_\odot]$       & 30.7 \\

$\beta$                              & 1.0 &
$\sigma_{g,\rm high}\,[M_\odot]$    & 6.3 \\

$m_{\rm min}\,[M_\odot]$             & 4.6 &
$\lambda_g$                          & 0.4 \\

$m_{\rm max}\,[M_\odot]$             & 86.3 &
$\lambda_{g,\rm low}$                & 0.8 \\

$\delta_m\,[M_\odot]$                & 4.8 &
$\gamma$                             & 3.3 \\

$\mu_{g,\rm low}\,[M_\odot]$         & 9.7 &
$\kappa$                             & 2.9 \\

$\sigma_{g,\rm low}\,[M_\odot]$      & 0.7 &
$z_p$                                & 2.5 \\

$H_0\,[{\rm km\,s^{-1}\,Mpc^{-1}}]$  & 67.4 &
$\Omega_m$                           & 0.315 \\
\hline
\end{tabular}
\caption{Injected hyperparameter values for the simulated gravitational-wave event catalog used in this work. The BBH population parameters are taken from the maximum-likelihood values reported in~\cite{gwtc4cosmo}, while the cosmological parameter values are those from Planck18~\cite{planck2018}.}
\label{tab:hyperparams}

\end{table}
\par
\textit{Real Data}: 
To demonstrate agreement between the current and previous versions of \gwcosmo we also include analysis using \gwtc data. Considering a selection threshold of IFAR $<\,4\,\mathrm{yr^{-1}}$, consistent with that in~\cite{gwtc4cosmo}, we use 141 CBCs in a dark siren analysis with the \fullpop mass model and the \gladeplus $K$-band galaxy catalog with weighting proportional to galaxy luminosity. Additionally to evaluate our selection function we use the publicly available injections campaign~\cite{essick2025}. 
\section{KL Validation}\label{app:kl}
\begin{figure}
    \centering
    \includegraphics[width=\linewidth]{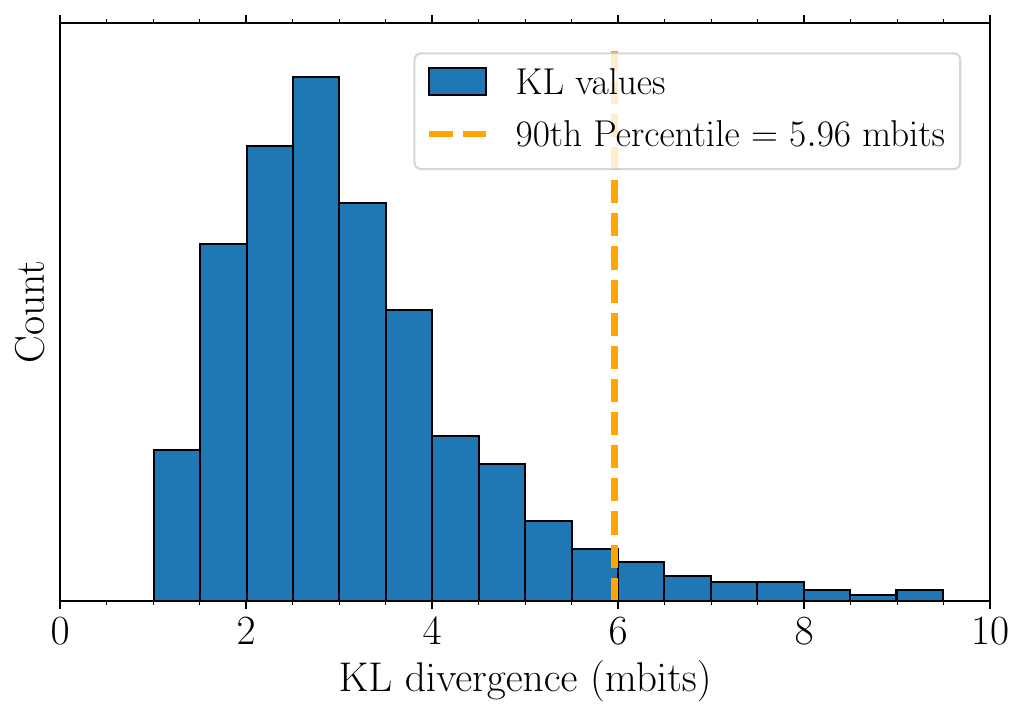}
    \caption{KL divergences calculated from two identical normal distributions, using $N_\mathrm{samples}=7000$ in order to show the scatter from using a finite number of samples. This was calculated $1000$ times, and the $90th$ percentile, denoted by the orange dashed line, was found to be $5.96\,\mathrm{mbits}$ even when comparing identical distributions.}
    \label{fig:kl_toy_model}
\end{figure}

In order to determine a threshold to decide on agreement between single marginal posteriors with the KL divergence, we employ a toy model with two identical normal distributions. By drawing $7000$ samples from both distributions (representative of the number of posterior samples produced in our analyses) and calculating the KL divergence for 1000 realisations. We find that $90\%$ of the values fall below is $5.96\, \mathrm{mbits}$, which can be seen in Fig.~\ref{fig:kl_toy_model}. This is similar to methods described in~\cite{romeroshaw2020, ashtontalbot2021} using another similar metric, the Jensen-Shannon divergence~\cite{jsd}. Given this, we elect to consider posteriors with values in or below this range to be in statistical agreement with one another, accounting for a finite number of samples. In the case of $N$-dimensional posteriors, we multiply this value by $N$ in order to assess agreement. 
\par 
Secondly, in order to quantify agreement between GPU and CPU posterior implementations in many dimensions, we establish a threshold on the noise from finite samples using symmetric resampling. We generate a null distribution by resampling the CPU posterior 1000 times - at each iteration, two independent half-sample subsets are drawn without replacement and the marginal KL divergence is computed for each of the 23 parameters using histogram-based estimation. The maximum of the marginal KL values across all parameters is calculated at each iteration. The 90th percentile of this distribution defines the noise threshold used to determine agreement. To assess the GPU posterior, we draw half-sample subsets from both CPU and GPU posteriors and compute the 90th percentile of marginal KL divergences as our test statistic. The GPU posterior is considered in agreement with the CPU posterior if this test statistic falls below the noise tolerance bound.
\section{Device Comparison and Energy Considerations}\label{app:devices}
Given that our analyses make use of a top-of-the-range GPU in the Nvidia H100 (Hopper architecture), we also compared this to the performance when using a more affordable device. This was chosen to be the Nvidia L40s (Ada Lovelace architecture), a device with a smaller overall memory and lower memory bandwidth, but improved single-precision optimisation. While still reporting a significant improvement in performance over the legacy CPU implementation, the L40s evaluates the likelihood a factor of $\sim4$ slower than the H100 when $N_\mathrm{events}=2000$. This points to the overall bottleneck on this device being memory bandwidth.
\begin{figure}[h]
    \centering
    \includegraphics[width=\linewidth]{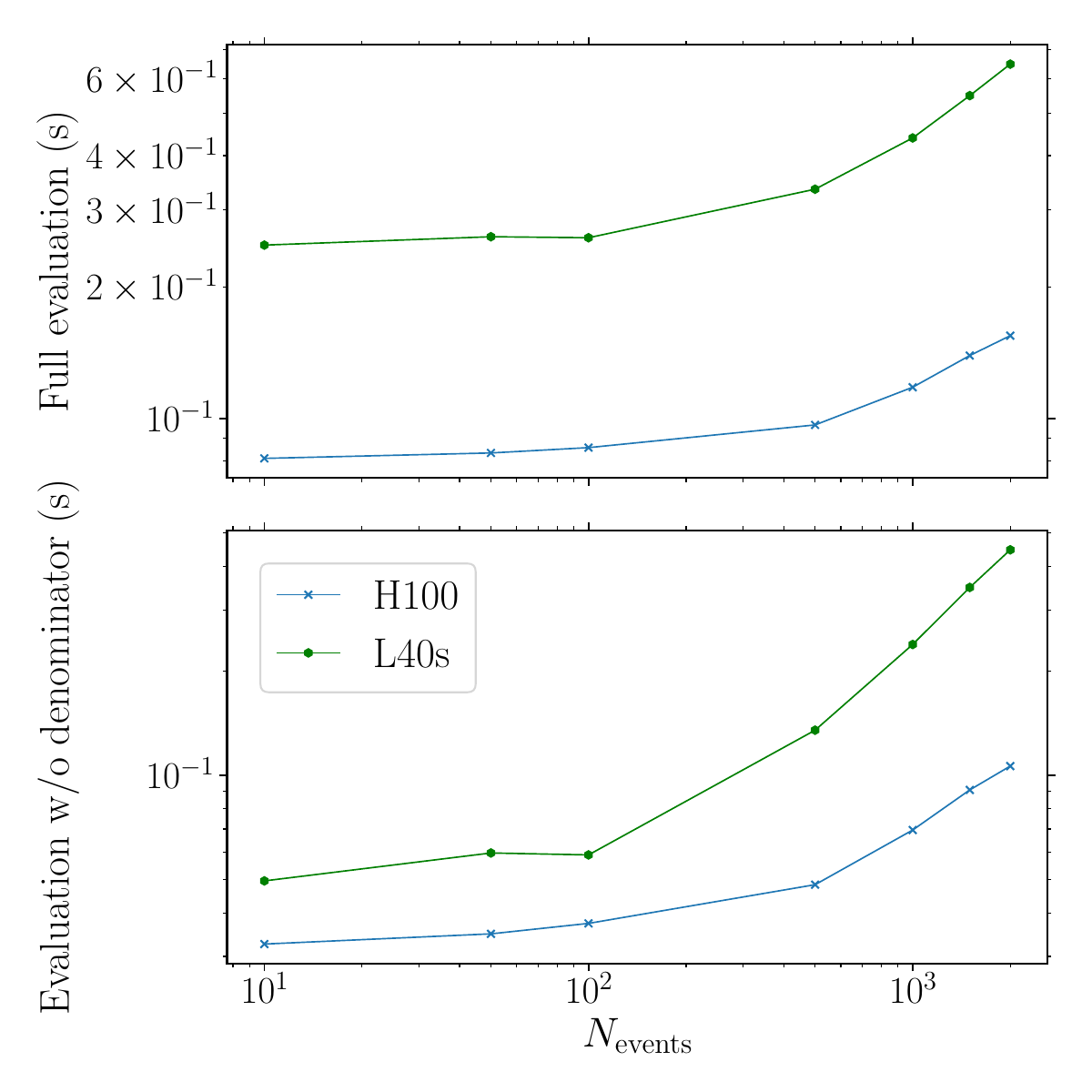}
    \caption{Bayesian inference wall-clock time comparison using difference devices. The NVIDIA H100 outperforms the more affordable L40S by a factor of 4 when using a full catalog of mock GW events. The performance gap is primarily attributed to the H100's superior memory bandwidth, making the top-of-the-range hardware more efficient for memory-bound likelihood calculations required in cosmological parameter estimation.}
    \label{fig:gpuscompare}
\end{figure}
\par In Table \ref{tab:energy}, we compare the energy consumption required to complete an analysis of \gwtc using 141 CBCs and the \fullpop mass model with the legacy CPU and updated GPU implementation. Using figures from \cite{ODYSSEEMURE2023}, the mean daily energy consumption for a residence in the European Union is taken to be $38.2 \,\mathrm{kWh}$, which is used to calculate the days equivalent row. The GPU implementation requires an order of magnitude less energy to complete this analysis when compared to it's CPU-bound counterpart.
\begin{table}[b]
\centering
\begin{tabular}{lcc}
\hline
\textbf{Metric} & \textbf{CPU (32 cores)} & \textbf{H100 GPU} \\
\hline
Wall time & 32 days (768 hours) & 15 hours \\
Compute time & 24{,}576 core-hours & 15 GPU-hours \\
Energy consumption & 230.4 kWh & 10.5 kWh \\
Days equivalent & $\approx6$ & $\approx 0.27$ \\
\hline
\end{tabular}
\caption{Comparison of computational cost, energy consumption, and estimated carbon emissions between a 32-core CPU implementation and an H100 GPU implementation with a maximum of 2500 samples allowed per sky pixel. The wall time is that of completing one analysis of \gwtc with the \fullpop mass model, the same as those shown in \S~\ref{sec:results}.}
\label{tab:energy}
\end{table}
\section{Bandwidth Estimators}\label{app:ests}
Scott’s rule provides an optimal bandwidth ($h$) for kernel density estimation, assuming the underlying data distribution is Gaussian~\cite{scott}. While computationally efficient, it relies heavily on the sample standard deviation, $\hat{\sigma}$, making it sensitive to outliers which can artificially inflate the bandwidth and oversmooth the density estimate. It takes the form
\begin{equation}
h = 1.06 \hat{\sigma} n^{-1/5}.
\end{equation}
Silverman’s rule improves upon Scott’s by introducing a more robust measure of spread~\cite{silverman}. By taking the minimum of the standard deviation and the scaled Interquartile Range (IQR), it prevents extreme outliers from excessively widening the bandwidth. It is more often used in cases where the distribution deviates from Gaussianity,
\begin{equation}
h = 0.9 \min \left( \hat{\sigma}, \frac{\text{IQR}}{1.34} \right) n^{-1/5}.
\end{equation}
The biweight midvariance is a robust estimator of scale~\cite{biweight}. Unlike the previous rules, it employs an indicator function $I(\mid u_i\mid<1)$ that acts as a ``hard limit" assigning zero weight to any observations falling more than 9 median absolute deviations (MAD) from the centre. This makes it significantly more resistant to heavy-tailed distributions than either Scott’s or Silverman’s methods. It is implemented as
\begin{equation}
h = \frac{n \sum_{i=1}^{n} (x_i - Q)^2 (1 - u_i^2)^4 I(|u_i| < 1)}{\left( \sum_{i} (1 - u_i^2)(1 - 5u_i^2) I(|u_i| < 1) \right)^2}.
\end{equation}

where the auxiliary variable $u_i$ is standardized by the median, $Q$,and the MAD,
\begin{equation}
u_i = \frac{x_i - Q}{9 \cdot \text{MAD}}.
\end{equation}

\bibliography{apssamp}

\end{document}